\newif\ifwordcount
\wordcountfalse

\documentclass[
reprint,
superscriptaddress,
showpacs,
nofootinbib,
amsmath,amssymb,
aps,
prl, % prl does not allow section reference. 
floatfix,
twocolumn,
]{revtex4-1}

\usepackage{graphicx}% Include figure files
\graphicspath{{./}}
\usepackage{dcolumn}% Align table columns on decimal point
\usepackage{bm}% bold math
\usepackage{xcolor}
\usepackage[colorlinks=true,allcolors=blue,linkcolor=blue,citecolor=blue]{hyperref}
\usepackage{amsmath, amsbsy}
\usepackage{amsthm}
\usepackage{amssymb}

\usepackage{tabularx}
\usepackage{xspace}
\usepackage{listings}
\usepackage{float}
\usepackage{booktabs}

\newcommand{\mnras}{Mon. Not. R. Astron. Soc.}
\newcommand{\aap}{Astron. Astrophys.}

\newcommand{\jcap}{J. Cosmol. Astropart. Phys.}
\newcommand{\procspie}{Proc. SPIE Int. Soc. Opt. Eng.}
\newcommand{\sovast}{Sov. Astron.}
\newcommand{\araa}{Annu. Rev. Astron. Astrophys.}

\newcommand{\patchun}[0]{\textrm{RA23}\xspace}
\newcommand{\patchdeux}[0]{\textrm{RA12}\xspace}
\newcommand{\patchtrois}[0]{\textrm{RA4.5}\xspace}
\newcommand{\pb}{\textsc{POLARBEAR}\xspace}

\newcommand{\vecL}[0]{ {\boldsymbol{L}} }
\newcommand{\vecl}[0]{ {\boldsymbol{\ell}} }
\newcommand{\vecx}[0]{ {\boldsymbol{x}} }
\newcommand{\cbb}{D_{\ell}^{\rm BB}}
\newcommand{\clbb}{C_{\ell}^{\rm BB}}
\newcommand{\ddbb}{\Delta \hat D_{\ell_b}^{\rm BB}\xspace}
\newcommand{\delensbias}{\Delta\hat D_{\ell_b}^{\rm BB, bias}\xspace}
\newcommand{\dabb}{\Delta A^{\rm BB}\xspace}
\newcommand{\dadelens}{\Delta A^{\rm delens}\xspace}
\newcommand{\commander}{\lstinline!COMMANDER!\xspace}
\newcommand{\hn}{\hat n}

\newcommand{\uK}{$\mu$K}
\newcommand{\farcm}{\mbox{\ensuremath{.\mkern-4mu^\prime}}}
\newcommand{\av}[1]{\left\langle #1 \right\rangle}

\newcommand{\doparagraph}[1]{\paragraph{#1.\!~{$-$}\!}}

\begin{document}
%\preprint{}
\title{Internal Delensing of Cosmic Microwave Background Polarization $B$-Modes with the \pb Experiment}
\author{S.~Adachi}
\affiliation{Department of Physics, Kyoto University, Kyoto 606-8502, Japan}
\author{M.~A.~O.~Aguilar~Fa\'undez}
\affiliation{Department of Physics and Astronomy, Johns Hopkins University, Baltimore, Maryland 21218, USA}
\affiliation{Departamento de F\'isica, FCFM, Universidad de Chile, Blanco Encalada 2008, Santiago, Chile}
\author{Y.~Akiba}
\affiliation{SOKENDAI (The Graduate University for Advanced Studies), Shonan Village, Hayama, Kanagawa 240-0193, Japan}
\author{A.~Ali}
\affiliation{Department of Physics, University of California, Berkeley, California 94720, USA}
\author{K.~Arnold}
\affiliation{Department of Physics, University of California, San Diego, California 92093-0424, USA}
\author{C.~Baccigalupi}
\affiliation{International School for Advanced Studies (SISSA), Via Bonomea 265, 34136, Trieste, Italy}
\affiliation{Institute for Fundamental Physics of the Universe (IFPU), Via Beirut 2, 34014 Trieste, Italy}
\affiliation{National Institute for Nuclear Physics (INFN), via Valerio 2, 34127 Trieste, Italy}
\author{D.~Barron}
\affiliation{Department of Physics and Astronomy, University of New Mexico, Albuquerque, New Mexico 87131, USA}
\author{D.~Beck}
\affiliation{AstroParticule et Cosmologie (APC), Univ Paris Diderot, CNRS/IN2P3, CEA/Irfu, Obs de Paris, Sorbonne Paris Cit\'e, France}
\author{F.~Bianchini}
\affiliation{School of Physics, University of Melbourne, Parkville, VIC 3010, Australia}
\author{J.~Borrill}
\affiliation{Computational Cosmology Center, Lawrence Berkeley National Laboratory, Berkeley, California 94720, USA}
\affiliation{Space Sciences Laboratory, University of California, Berkeley, California 94720, USA}
\author{J.~Carron}
\email[Corresponding author. ]{J.Carron@sussex.ac.uk}
\affiliation{Department of Physics \& Astronomy, University of Sussex, Brighton BN1 9QH, United Kingdom}
\author{K.~Cheung}
\affiliation{Department of Physics, University of California, Berkeley, California 94720, USA}
\author{Y.~Chinone}
\affiliation{Department of Physics, University of California, Berkeley, California 94720, USA}
\affiliation{Kavli Institute for the Physics and Mathematics of the Universe (Kavli IPMU, WPI), UTIAS, The University of Tokyo, Kashiwa, Chiba 277-8583, Japan}
\affiliation{Kavli Institute for the Physics and Mathematics of the Universe (WPI), Berkeley Satellite, the University of California, Berkeley, California 94720, USA}
\author{K.~Crowley}
\affiliation{Department of Physics, University of California, Berkeley, California 94720, USA}
\author{H.~El~Bouhargani}
\affiliation{AstroParticule et Cosmologie (APC), Univ Paris Diderot, CNRS/IN2P3, CEA/Irfu, Obs de Paris, Sorbonne Paris Cit\'e, France}
\author{T.~Elleflot}
\affiliation{Department of Physics, University of California, San Diego, California 92093-0424, USA}
\author{J.~Errard}
\affiliation{AstroParticule et Cosmologie (APC), Univ Paris Diderot, CNRS/IN2P3, CEA/Irfu, Obs de Paris, Sorbonne Paris Cit\'e, France}
\author{G.~Fabbian}
\email[Corresponding author. ]{G.Fabbian@sussex.ac.uk}
\affiliation{Department of Physics \& Astronomy, University of Sussex, Brighton BN1 9QH, United Kingdom}
\author{C.~Feng}
\affiliation{Department of Physics, University of Illinois at Urbana-Champaign, 1110 West Green Sreet, Urbana, Illinois, 61801, USA}
\author{T.~Fujino}
\affiliation{Yokohama National University, Yokohama, Kanagawa 240-8501, Japan}
\author{N.~Goeckner-Wald}
\affiliation{Department of Physics, University of California, Berkeley, California 94720, USA}
\author{M.~Hasegawa}
\affiliation{High Energy Accelerator Research Organization (KEK), Tsukuba, Ibaraki 305-0801, Japan}
\author{M.~Hazumi}
\affiliation{High Energy Accelerator Research Organization (KEK), Tsukuba, Ibaraki 305-0801, Japan}
\affiliation{Kavli Institute for the Physics and Mathematics of the Universe (Kavli IPMU, WPI), UTIAS, The University of Tokyo, Kashiwa, Chiba 277-8583, Japan}
\affiliation{Institute of Space and Astronautical Science (ISAS), Japan Aerospace Exploration Agency (JAXA), Sagamihara, Kanagawa 252-0222, Japan}
\affiliation{SOKENDAI (The Graduate University for Advanced Studies), Shonan Village, Hayama, Kanagawa 240-0193, Japan}
\author{C.~A.~Hill}
\affiliation{Department of Physics, University of California, Berkeley, California 94720, USA}
\author{L.~Howe}
\affiliation{Department of Physics, University of California, San Diego, California 92093-0424, USA}
\author{N.~Katayama}
\affiliation{Kavli Institute for the Physics and Mathematics of the Universe (Kavli IPMU, WPI), UTIAS, The University of Tokyo, Kashiwa, Chiba 277-8583, Japan}
\author{B.~Keating}
\affiliation{Department of Physics, University of California, San Diego, California 92093-0424, USA}
\author{S.~Kikuchi}
\affiliation{Yokohama National University, Yokohama, Kanagawa 240-8501, Japan}
\author{A.~Kusaka}
\affiliation{Physics Division, Lawrence Berkeley National Laboratory, Berkeley, California 94720, USA}
\affiliation{Department of Physics, The University of Tokyo, Tokyo 113-0033, Japan}
\affiliation{Kavli Institute for the Physics and Mathematics of the Universe (WPI), Berkeley Satellite, the University of California, Berkeley, California 94720, USA}
\affiliation{Research Center for the Early Universe, School of Science, The University of Tokyo, Tokyo 113-0033, Japan}
\author{A.~T.~Lee}
\affiliation{Department of Physics, University of California, Berkeley, California 94720, USA}
\affiliation{Physics Division, Lawrence Berkeley National Laboratory, Berkeley, California 94720, USA}
\affiliation{Radio Astronomy Laboratory, University of California, Berkeley, California 94720, USA}
\author{D.~Leon}
\affiliation{Department of Physics, University of California, San Diego, California 92093-0424, USA}
\author{E.~Linder}
\affiliation{Space Sciences Laboratory, University of California, Berkeley, California 94720, USA}
\author{L.~N.~Lowry}
\affiliation{Department of Physics, University of California, San Diego, California 92093-0424, USA}
\author{F.~Matsuda}
\affiliation{Kavli Institute for the Physics and Mathematics of the Universe (Kavli IPMU, WPI), UTIAS, The University of Tokyo, Kashiwa, Chiba 277-8583, Japan}
\author{T.~Matsumura}
\affiliation{Kavli Institute for the Physics and Mathematics of the Universe (Kavli IPMU, WPI), UTIAS, The University of Tokyo, Kashiwa, Chiba 277-8583, Japan}
\author{Y.~Minami}
\affiliation{High Energy Accelerator Research Organization (KEK), Tsukuba, Ibaraki 305-0801, Japan}
\author{T.~Namikawa}
\affiliation{DAMTP, University of Cambridge, Cambridge CB3 0WA, United Kingdom}
\author{M.~Navaroli}
\affiliation{Department of Physics, University of California, San Diego, California 92093-0424, USA}
\author{H.~Nishino}
\affiliation{Research Center for the Early Universe, School of Science, The University of Tokyo, Tokyo 113-0033, Japan}
\author{J.~Peloton}
\affiliation{Laboratoire de l'Acc\'el\'erateur Lin\'eaire, Universit\'e Paris-Sud, CNRS/IN2P3, Orsay, France}
\author{A.~T.~P.~Pham}
\affiliation{School of Physics, University of Melbourne, Parkville, VIC 3010, Australia}
\author{D.~Poletti}
\affiliation{International School for Advanced Studies (SISSA), Via Bonomea 265, 34136, Trieste, Italy}
\affiliation{Institute for Fundamental Physics of the Universe (IFPU), Via Beirut 2, 34014 Trieste, Italy}
\affiliation{National Institute for Nuclear Physics (INFN), via Valerio 2, 34127 Trieste, Italy}
\author{G.~Puglisi}
\affiliation{Department of Physics, Stanford University, Stanford, CA, 94305}
\affiliation{Kavli Institute for Particle Astrophysics and Cosmology, SLAC National Accelerator Laboratory, 2575 Sand Hill Road, Menlo Park, California 94025}
\author{C.~L.~Reichardt}
\affiliation{School of Physics, University of Melbourne, Parkville, VIC 3010, Australia}
\author{Y.~Segawa}
\affiliation{SOKENDAI (The Graduate University for Advanced Studies), Shonan Village, Hayama, Kanagawa 240-0193, Japan}
\author{B.~D.~Sherwin}
\affiliation{Kavli Institute for Cosmology Cambridge, Cambridge CB3 OHA, United Kingdom}
\author{M.~Silva-Feaver}
\affiliation{Department of Physics, University of California, San Diego, California 92093-0424, USA}
\author{P.~Siritanasak}
\affiliation{Department of Physics, University of California, San Diego, California 92093-0424, USA}
\author{R.~Stompor}
\affiliation{AstroParticule et Cosmologie (APC), Univ Paris Diderot, CNRS/IN2P3, CEA/Irfu, Obs de Paris, Sorbonne Paris Cit\'e, France}
\author{O.~Tajima}
\affiliation{Department of Physics, Kyoto University, Kyoto 606-8502, Japan}
\author{S.~Takatori}
\affiliation{SOKENDAI (The Graduate University for Advanced Studies), Shonan Village, Hayama, Kanagawa 240-0193, Japan}
\affiliation{High Energy Accelerator Research Organization (KEK), Tsukuba, Ibaraki 305-0801, Japan}
\author{D.~Tanabe}
\affiliation{SOKENDAI (The Graduate University for Advanced Studies), Shonan Village, Hayama, Kanagawa 240-0193, Japan}
\affiliation{High Energy Accelerator Research Organization (KEK), Tsukuba, Ibaraki 305-0801, Japan}
\author{G.~P.~Teply}
\affiliation{Department of Physics, University of California, San Diego, California 92093-0424, USA}
\author{C.~Verg\`es}
\affiliation{AstroParticule et Cosmologie (APC), Univ Paris Diderot, CNRS/IN2P3, CEA/Irfu, Obs de Paris, Sorbonne Paris Cit\'e, France}
\collaboration{The \textsc{POLARBEAR} collaboration}
\noaffiliation

%\pacs{}
\date{\today}

\begin{abstract}
Using only cosmic microwave background polarization data from the \textsc{POLARBEAR} experiment, we measure $B$-mode polarization delensing on subdegree scales at more than $5\sigma$ significance. We achieve a 14\% $B$-mode power variance reduction, the highest to date for internal delensing, and improve this result to 22\% by applying for the first time an iterative \emph{maximum a posteriori} delensing method. Our analysis demonstrates the capability of internal delensing as a means of improving constraints on inflationary models, paving the way for the optimal analysis of next-generation primordial $B$-mode experiments.
\end{abstract} 

\ifwordcount
\else
\maketitle
\fi

%-------Introduction
\doparagraph{Introduction}
Inflation is a paradigm which can explain the physics of the primordial Universe. It features an early epoch of accelerated expansion during which the primordial density perturbations as well as a generic stochastic background of gravitational waves are produced. The latter subsequently imprints a unique signature in the anisotropies of the cosmic microwave background (CMB) polarization, curl-like patterns ($B$-modes), most prominent on degree angular scales~\cite{seljak-zaldarriaga-gw, kks1997, polnarev1985}. The amplitude of such a signal (usually parametrized by the tensor-to-scalar ratio $r$) can be related to the energy scale at which inflation took place and thus is one of the most promising probes of the physics of the early Universe~\cite{kamionkowski2016}. However, large-scale structures (LSS) in the Universe distort the predominant gradient-like E-modes of CMB polarization (that are mainly generated by the primordial scalar perturbations) through weak gravitational lensing, creating additional $B$-mode polarization~\cite{Lewis:2006fu, seljak2004} that contaminates the tensor signal.

The lensing $B$-modes act as a source of variance, and will soon limit primordial $B$-mode searches. Removing the lensing effects in CMB maps (delensing) will become a necessary data analysis step~\cite{kesden2002}. Delensing requires the subtraction of a template of the lensing $B$-mode signal constructed from observed E-modes and a tracer of the mass distribution that lensed the CMB. This tracer can be obtained from CMB through its lensing potential (internal delensing) or using external astrophysical data. 
Delensing has been demonstrated on data only recently~\cite{Larsen:2016wpa,Carron:2017vfg, Manzotti:2017net, Aghanim:2018oex}. A maximal reduction in B-power of 28\% has been achieved using the cosmic infrared background as the lensing tracer~\cite{Manzotti:2017net, Sherwin:2015baa}. The only internal delensing attempts so far used Planck data and achieved a 5\%--7\% reduction in power limited by the high noise in the tracer measurement~\cite{Carron:2017vfg, Aghanim:2018oex}. While CIB and LSS delensing will remain more powerful in the next few years, internal delensing is expected to eventually become more effective and remove the lensing $B$-modes almost optimally~\cite{Carron:2018lcr} when suitably low-noise data are available~\cite{Hirata:2002jy}.

We report here a delensing analysis of the subdegree $B$-mode signal angular power spectrum $\clbb$ of the CMB polarization experiment \pb~\cite{arnold2012, kermish2012}. We test two types of internal lensing estimators: the standard quadratic estimator (QE) $\hat\phi^{\rm QE}$~\cite{Hu:2001kj} and a more powerful \emph{maximum a posteriori} (MAP) iterative method $\hat\phi^{\rm MAP}$~\cite{Hirata:2002jy, Carron:2017mqf}, applied here to data for the first time.

The reconstruction noise of CMB internal estimates originates from random anisotropic features in the CMB maps that were interpreted as lensing. Hence, an attempt to remove the lensing features using these tracers can suppress too much anisotropy of the CMB maps. Large delensinglike signatures (called internal delensing bias), unrelated to actual delensing, can then be found in the delensed CMB spectra~\cite{Teng:2011xc, Carron:2017vfg, Sehgal:2016eag}. 
To mitigate this problem we introduce a dedicated technique applicable both to the QE and MAP estimations.

%-------Data and simulations
\doparagraph{Data and simulations}
 We use the first two seasons of observations between 2012 and 2014, covering an effective sky area of 25 deg$^2$ at $3\farcm5$ resolution distributed over three sky patches chosen for their low foreground contamination, referred to as RA23, RA12 and RA4.5. The effective white-noise levels in the full-season coadded map of the Stokes parameters $Q^{\rm dat}$ and $U^{\rm dat}$ reach 6, 7, and 10\,\uK arcmin respectively. These are among the deepest observations of CMB polarization to date at high angular resolution. This dataset is well suited for an internal delensing analysis as it provides good signal-to-noise measurements of both the lensing tracer and CMB polarization. 
Details of the \pb data analysis are given in Refs.~\cite{Ade:2014afa} (PB14) and~\cite{Ade:2017uvt} (PB17). In this work we assume Planck 2015~\cite{Ade:2015xua} as our fiducial $\Lambda$CDM cosmology and use CMB maps produced with \pb pipeline A. We correct the maps for the absolute calibration, polarization efficiency and polarization angle miscalibration following PB17 before any further processing. We use Fourier modes $500 \leq \ell \leq 2500$ to construct the lensing tracers and report delensing results in four linearly spaced multipole bins between $500 \leq \ell \leq 2100$.
To characterize uncertainties in our analysis we use two sets of 500 simulated \pb datasets including realistic noise and data processing effects as in PB17. The two sets share the same noise realizations but use lensed or Gaussian CMB drawn from a lensed CMB power spectrum as sky signal. We refer to these sets of simulations as non-Gaussian and Gaussian simulations respectively.

%-------Power spectrum estimation
\doparagraph{Power spectrum estimation}
Following PB14 and PB17, we estimate the $E$- and $B$-mode power spectra~\cite{seljak-zaldarriaga1997} from the daily $Q$ and $U$ maps through an inverse noise variance weighted average of their pure-pseudo cross-spectra~\cite{smith2007, grain2009} accounting for the sky masking, telescope beam, and data processing effects~\cite{hivon2002}. To estimate the delensed spectra we follow the same pipeline, but first subtract the templates of the lensing $B$-mode described below from each daily map prior to the cross-spectrum calculation. We denote  the difference in power after and before delensing by $\Delta D_{\ell}^{\rm BB} \equiv D_{\ell}^{\rm BB, delens} - D_{\ell}^{\rm BB}$, where $\cbb=\ell(\ell+1)\clbb/2\pi$.

%-------QE lens estimate
\doparagraph{Quadratic estimate}
From the full-season-coadded maps $X^{\rm dat} = (Q^{\rm dat}, U^{\rm dat})$ we produce Wiener filtered $E$- and $B$-modes $X^{\rm WF}_\vecl\!\equiv\!(E^{\rm WF}_\vecl, B^{\rm WF}_\vecl)$ in the flat-sky approximation as follows. We build pixel-space diagonal noise covariance matrices $N$ from our noise simulations, which include inhomogeneities induced by the observing strategy. Combining this with the full effective PB17 transfer function $\mathcal B$ (defined as mapping the CMB $E$ and $B$ Fourier modes to pixelized Stokes data, including the instrument beam and processing transfer function), we have
\ifwordcount
\else
\begin{equation}\label{eq:filter}
	X^{\rm WF}_\vecl\!\equiv\!\left[\!\begin{pmatrix}\!\frac{1}{C_\ell^{\rm EE}}&0\\0&\frac{1}{C_\ell^{\rm BB}}\!\end{pmatrix}\!\delta_{\vecl\vecl'}\!+\!\left[\mathcal B^\dagger N^{-1}\mathcal B\right]_{\vecl\vecl'}\!\right]^{-1}\hspace{-.025\textwidth}\mathcal B^\dagger\!N^{-1}\!X^{\rm dat}\!.
\end{equation}
\fi
This neglects the small $E$ to $B$ leakage caused by data processing as well as anisotropies in the transfer function. Both effects are included in the simulations and only result in a slight suboptimality of the lensing tracer.
We mask pixels with estimated noise level larger than 55\,\uK arcmin, and include PB17 point source masks. To reduce the internal delensing biases, we modify the $N$ matrix by artificially assigning extra noise $\sigma_b$ to every single $B$-mode within the multipole bin $b$ that we try to delens. Such modes are the main contributors to the biases. We refer to this procedure as the overlapping $B$-modes deprojection (OBD). The $N^{-1}$ matrix is then replaced by the $N^{-1}_{(b)}$ matrix
\ifwordcount
\else
\begin{equation}\label{eq:projector}
N^{-1}_{(b)} \equiv N^{-1} - N^{-1}\mathcal T_b\left( \frac 1 {\sigma_b^2} + \mathcal T^\dagger_b N^{-1} \mathcal T_b\right)^{-1}\mathcal T^\dagger_b N^{-1},
\end{equation}
\fi
where for every $B$-mode multipole $\vecl_B = \ell_Be^{i \phi_{\vecl_B}}$ within a multipole bin $\left[\mathcal T_b\right]_{Q(\vecx_i)\vecl_B } =   \sin 2\phi_{\vecl_B} e^{i\vecl_B \cdot \vecx_i}$ and $\left[\mathcal T_b\right]_{ U(\vecx_i)\vecl_B} = \cos 2\phi_{\vecl_B} e^{i\vecl_B \cdot \vecx_i}$.
The complete masking of these modes is achieved only for infinite $\sigma_b^2$, but in this case the  inversion of the bracketed matrix in Eq.~\eqref{eq:projector} becomes numerically unstable. To avoid this,  we chose a high, but finite, noise amplitude $\sigma_b  = 1000$\uK arcmin to sufficiently down weight them. Using $\sigma_b  = 100$\uK arcmin does not change our results. Equation~\eqref{eq:filter} is then evaluated with a simple conjugate gradient solver.
From these filtered maps an unnormalized quadratic estimate of the CMB lensing Fourier modes $\hat g_\vecL$ is built following Ref.~\cite{Carron:2017mqf}, using the minimum variance combination of the EE and EB estimators (in the fiducial model). At the \pb level of sensitivity, the polarization data provide a CMB lensing reconstruction noise lower than that achievable using temperature data on all angular scales. The EB estimator in particular has the lowest noise in RA23 and RA12 sky patches. The estimate is then normalized and Wiener filtered as $\hat \phi_{\vecL}^{\rm QE, WF} = \epsilon_L N^{(0)}_L\left(\hat g_\vecL  -\left \langle \hat g_\vecL \right\rangle_{\rm MC}\right)$, where $\epsilon_L\equiv C_L^{\phi\phi, \rm fid}/(C_L^{\phi\phi, \rm fid} + N_L^{(0)})$, $N_L^{(0)}$ is the QE reconstruction noise level~\cite{Okamoto:2003zw} as predicted from the central noise levels of the patches, their effective transfer functions, CMB $E$ and $B$ multipole cuts. $C_L^{\phi\phi, \rm fid}$ is our fiducial lensing potential power spectrum. This isotropic normalization is adequate in the patch centers where delensing is most efficient, but results in a slight down weighting of the tracer toward the edges where the noise is higher. Finally, $\left \langle \hat g_\vecL \right\rangle_{\rm MC}$ is the ``mean-field'' used to subtract sources of anisotropies unrelated to lensing~\citep{Hanson:2009gu} obtained by averaging 200 simulations. $\epsilon_L$ may be interpreted as a naive estimator of the scale-dependent delensing efficiency in the patch centers~\cite{Sherwin:2015baa}.
OBD trades delensing efficiency for lower delensing biases. In RA23 this reduces $\epsilon_L$ by $\sim35\%$, $20\%$, $10\%$, and $5\%$ for our four bins, compared to no deprojection. This issue is less severe for experiments aiming at delensing degree-scale $B$-modes as in this case the modes to exclude are restricted to the largest scales, which carry little information for the lensing potential reconstruction.

%-------MAP lens estimate
\doparagraph{Iterative estimate}
 The construction of the MAP lensing estimate follows closely Ref.~\cite{Carron:2017mqf} (with the addition of OBD), which can be briefly summarized as follows: at each iteration step, the filter in Equation~\eqref{eq:filter} is replaced with a similar filter with vanishing $C_\ell^{\rm BB}$ but which includes the lensing deflection estimate in $\mathcal{B}$.
  This reconstructs a partially delensed CMB. Then, a quadratic estimator with modified weights corrected by a mean-field term is used to capture residual lensing from these new maps. Our treatment of the mean field is simpler than Ref.~\cite{Carron:2017mqf}. The mean field is small at the scales of interest, and its reevaluation at each step and band-power bin for each simulation realization is expensive. Thus, we use the same mean field computed for $\hat\phi^{QE}$ at all steps. We perform three iterations after which we see no significant improvement.

%-------B lensing template
\doparagraph{Lensing B-mode templates}
For each $\hat \phi_{\vecL}$ estimate, we build a $B$-mode template synthesizing first the $\hat \phi$ map, the polarization map $P^{(E^t)} = Q^{(E^t)} +i U^{(E^t)}$ from an $E$-mode template ($E^t$) and then projecting the remapped polarization template $P^{(E^t)}[\hn + \nabla  \hat\phi(\hn)]$ into $B$-modes ($B^t$). 
For $E^t$ we use the $E^{\rm WF}_\vecl$ solution of Eq.~\eqref{eq:filter} without any $B$-mode deprojection and apply the multipoles cuts $500 \le \ell_E \le 2500$. The excluded multipoles contribute 10\% of the lensing $B$-mode power in our lowest bin $\ell_B\sim500$~\cite{fabbian2013}, and  percent level at higher $\ell_B$. The impact on our delensing efficiency is thus minor. All lensing multipoles $L\leq 100$ are cut from the lensing map. This removes all scales where the mean-field is large compared to the signal, but does not affect the delensing capability of the tracer.

%-------Delensing biases
\doparagraph{Internal delensing bias}
\begin{figure}[!htbp]
\centering
\ifwordcount
\else
\includegraphics[width = 0.5\textwidth]{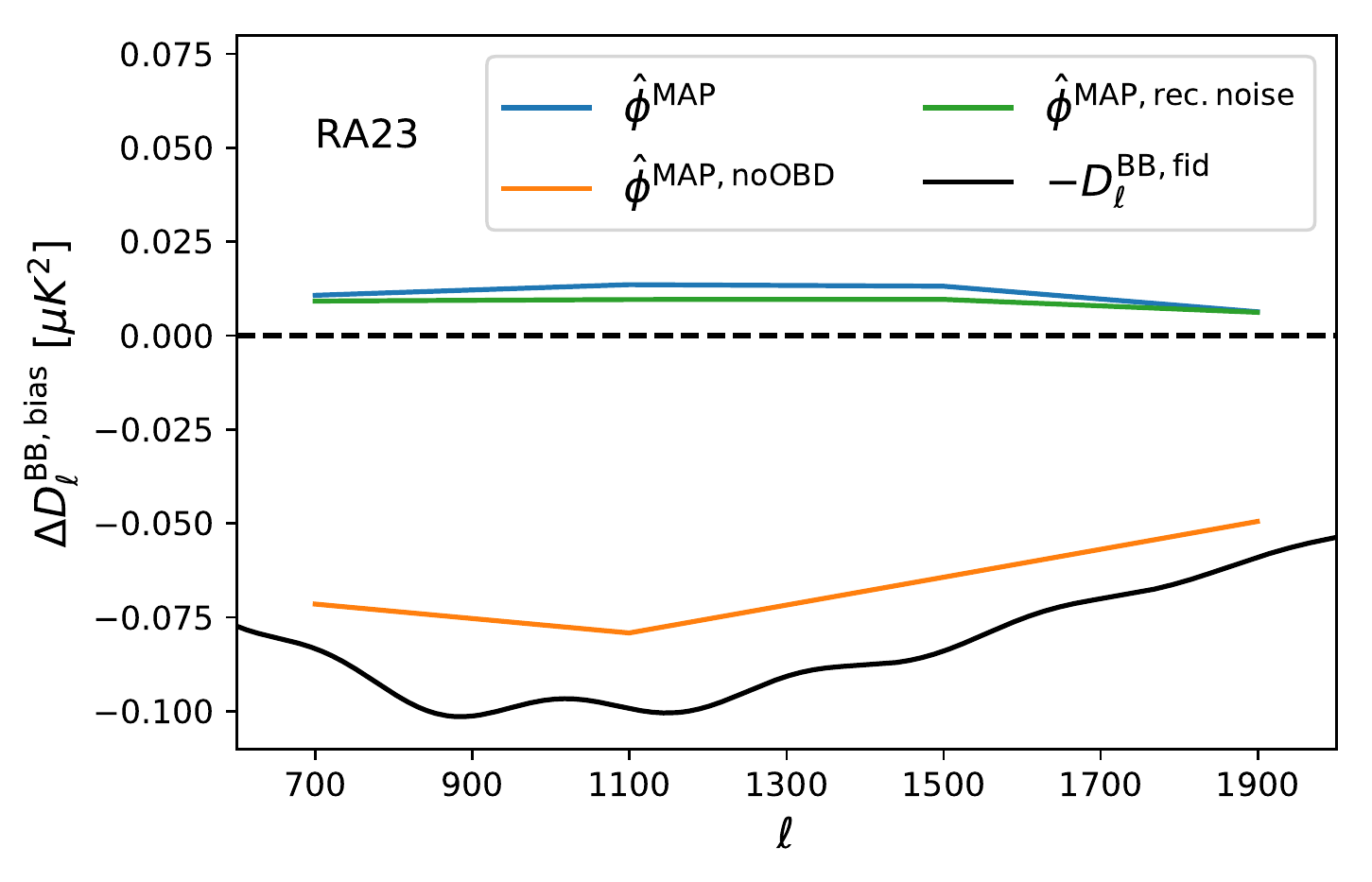}
\fi
 \caption{\label{fig:Gbias} Total delensing bias (defined as the result of the internal delensing operation applied to Gaussian CMB simulations) in our $\hat \phi^{\rm MAP}$ analysis (blue line) of RA23. The contribution due to the noise of $\hat \phi^{\rm MAP}$ acting on the $E$-mode template is shown in green. The delensing bias one would obtain without the overlapping $B$-modes deprojection (orange line) approaches the amplitude of $D_{\ell}^{\rm BB}$ of our fiducial cosmology (black line), mimicking perfect delensing.}
\end{figure}
$B^t$ is built out of three CMB fields: $E^t$, $E^{\rm WF}$, and $B^{\rm WF}$, where the last two are used to estimate $\hat \phi_{\vecL}$.  In a standard QE implementation the leading contribution to the internal delensing bias (though not all of it at low-noise levels~\cite{Namikawa:2017iak}) is sourced by the disconnected (Gaussian) correlation functions involving four CMB fields.
The leading contributing terms in the spectrum[$(B^{\rm dat}-B^t)^2$, schematically] have the form $[E^{\rm t} \star \hat \phi^{\rm noise}(E^{\rm WF}, B^{\rm WF})] \cdot B^{\rm dat}$, where $\star$ denotes the template building operation, the center dot $(\cdot)$ denotes the cross-spectrum between the template and the data, and $\hat\phi^{\rm noise}$ being the noise of the lensing tracer reconstructed using the EB estimator. 
Following Ref.~\cite{Carron:2017vfg}, we compute the delensing bias as $\Delta\hat D_{\ell_b}^{\rm BB, bias}\equiv\langle\Delta\hat D_{\ell_b}^{\rm BB}\rangle_G$, where $\langle\cdot\rangle_G$ denotes that the entire internal delensing operation is performed on Gaussian simulations, and averaged over. Since the simulations are Gaussian, the estimated lensing tracers are pure noise, and this term captures these disconnected correlators. In Fig.~\ref{fig:Gbias} we show the MAP $\delensbias$ for the RA23 data (the QE curves are similar). If no OBD is performed we see a strong negative signal similar to our negative fiducial $\cbb$, creating the illusion of an almost perfect delensing (orange line).  OBD prevents correlating overlapping modes in $B^{\rm WF}$  and $B^{\rm dat}$, reducing the entire bias by almost an order of magnitude (blue). 
Were the tracer noise statistically independent of the map being delensed, we would only see the (positive) B-power induced by the remapping of $P^{(E^t)}$ by the tracer noise (green). This contribution can be quantified by delensing each simulation realization with an independent MAP tracer. 
The dominant residual contribution to the delensing bias after OBD is mostly sourced by this term, owing to $B$-modes at $\ell_B > 2500$ leaking to lower $\ell_B$ in both $B^{\rm WF}$  and $B^{\rm dat}$ due to the presence of the mask which convolves different angular scales. We verified this directly with the help of another, simpler set of simulations where all modes above $\ell_B > 2500$ were artificially zeroed out prior the analysis.

\begin{figure}[!htbp]
\centering
\ifwordcount
\else
\includegraphics[width = 0.5\textwidth]{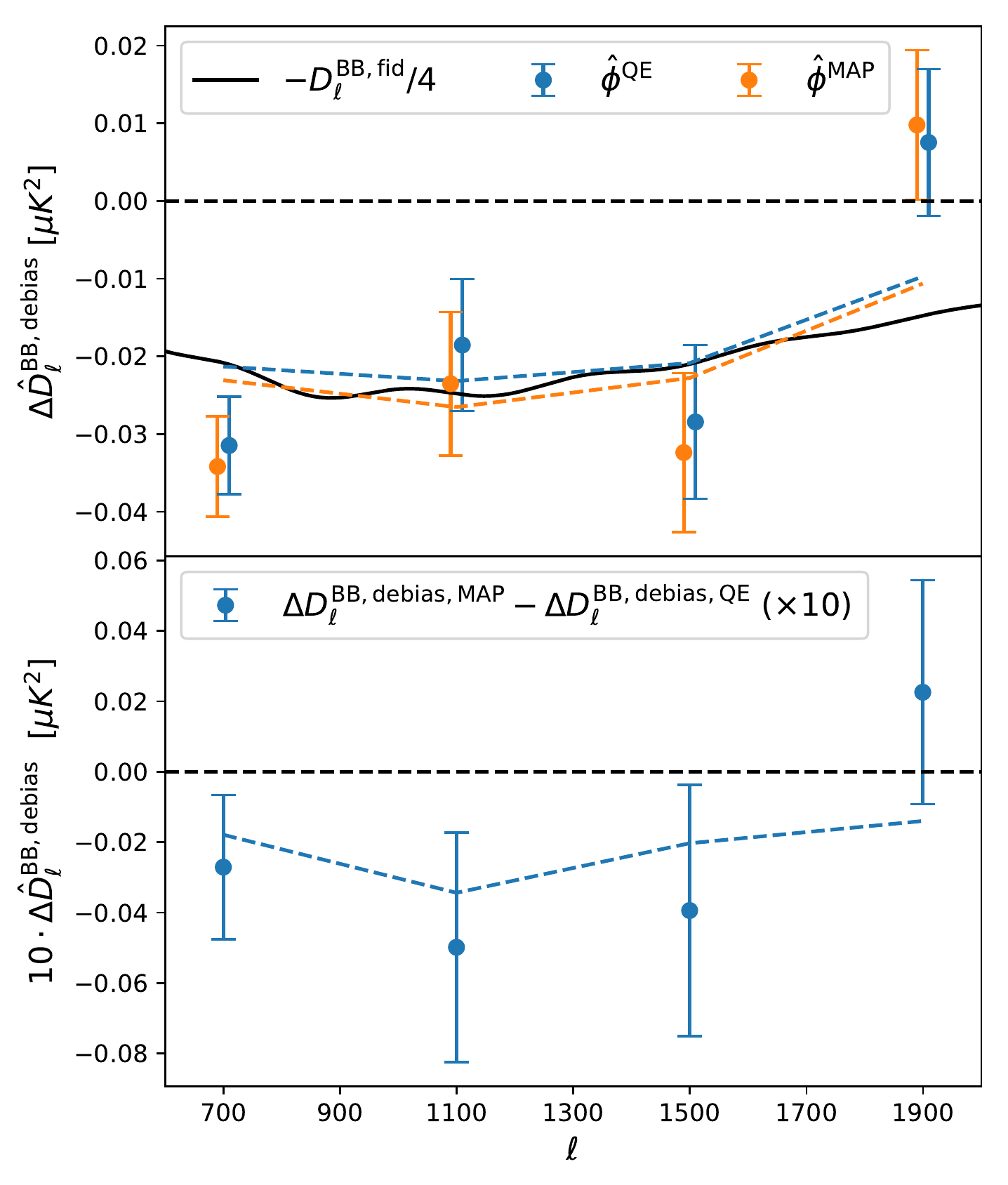}
\fi
 \caption{Top: Inverse-variance combination of the debiased spectral differences $\Delta \hat D_\ell^{\rm BB,debiased}$ measured in the \pb sky patches using QE (blue) and MAP (orange) delensing. The dashed lines show expectations obtained as average of results computed on simulations. Bottom: Difference between MAP and QE delensed $\Delta \hat D_\ell^{\rm BB,debiased}$ compared to simulation expectations (dashed lines). As both these quantities are highly correlated, the error bar of this statistic is significantly reduced when compared to those of the top panel. The significance of this difference being nonzero is $2.1\sigma$.}
 \label{fig:specdiff}
\end{figure}

%-------Results
\doparagraph{Results}
We build debiased band powers according to
\ifwordcount
\else
\begin{equation}\label{eq:DmDg}
	\Delta \hat D_{\ell_b}^{\rm BB, debiased}=\Delta \hat D_{\ell_b}^{\rm BB} -\delensbias.
\end{equation}
\fi
\noindent
In Fig.~\ref{fig:specdiff} we show the inverse-variance weighted combination of $\Delta \hat D_\ell^{\rm BB, debiased}$ in the \pb patches. Table~\ref{table:stats} shows the values of the amplitude $\dadelens$ of the simulation predictions of $\Delta \hat D_{\ell_b}^{\rm BB, debiased}$ fit to the data. By construction, these band-powers are in practice free of the internal delensing bias, and $\dadelens=0$ in the absence of lensing signatures in the data.
For the patch-combined measurement, we detect a nonzero $\dadelens$ with a significance $\dadelens/\sigma_{\dadelens}$ of $5\sigma$ using $\hat\phi^{\rm QE}$, consistent with simulation predictions ($\dadelens=1$). The significance of the patch-combined measurement increases to $5.3\sigma$ using $\hat\phi^{\rm MAP}$. Our deepest patch RA23 alone provides a $4\sigma$ measurement.  
\noindent
\begin{table}[h]
\ifwordcount
\else
\begin{ruledtabular}
\begin{tabular}{lcc}
$\Delta \hat D_{\ell_b}^{\rm BB, debiased}$& $\dadelens[\hat \phi^{\rm QE}]$ & $\dadelens[\hat \phi^{\rm MAP}]$\\
%----Numbers fitted to sim pred:
\hline
RA23 &1.26 $\pm$ 0.33 & 1.38 $\pm$ 0.32\\ 
RA12 &1.16 $\pm$ 0.39 & 1.09 $\pm$ 0.37\\ 
RA4.5 &0.79 $\pm$ 0.59 & 0.59 $\pm$ 0.57\\ 
Patch combined & 1.22 $\pm$ 0.24 & 1.24 $\pm$ 0.23\\ 
\end{tabular}
\end{ruledtabular}
\fi
\caption{\label{table:stats}Fit of the amplitude $\dadelens$ of the simulation prediction to the debiased delensed spectrum difference $\Delta \hat D_{\ell_b}^{\rm BB, debiased}$. Error bars were calculated using the Gaussian simulation set. A nonzero delensing signal is measured at more than $5\sigma$ after combining the data of all patches, consistent with simulation predictions $(\dadelens = 1)$.}
\end{table}
\noindent
In all cases, $\dadelens$ agrees with expectations from simulations (shown as dashed line in Fig.~\ref{fig:specdiff}), where the MAP delensing always outperforms QE. While MAP delensing does increase the significance of our results, we see evidence for the improvement over QE in the data only at modest significance. The difference of $\Delta \hat D_{\ell_b}^{\rm BB, debiased}$ computed with MAP and QE is nonzero at $2.1\sigma$ significance, consistent with simulation expectations. The fluctuations in this statistic are caused by the decoherence of the MAP-delensed and QE-delensed maps sourced by the slightly different noise components in the tracers. A deviation from zero of this difference is thus sourced by a difference in the delensed signal.

How much lensing $B$-mode power variance did we actually remove? The debiasing procedure subtracts $B$-power that acts as a source of additional variance in parameter inference. Hence, the relevant quantity is the reduction of power without any debiasing\footnote{This is not always the case for internal delensing performed at the degree-scale, where both the residual power and variance carry a strong $r$-dependence that has to be carefully characterized~\cite{Namikawa:2015tba, Carron:2018lcr}). Our bias is sourced by high-$\ell_B$ noise with no cosmological dependence.}. We find a reduction of $B$-power of 14\% ($\hat\phi^{\rm QE}$) and 22\% ($\hat\phi^{\rm MAP}$) for our deepest patch RA23, in agreement with simulation expectations [$(13 \pm 9)\%$ and $ (15 \pm 9)\%$, respectively, for its mean value]. It is more difficult to distinguish the QE from the MAP result without debiasing on real data or on a single realization of the simulations. The observed difference between MAP and QE is thus measured at only $1.1\sigma$, down from $2.1\sigma$ when performing debiasing. 
For MAP, RA12 and RA4.5 achieved a 15\% and 1\% power reduction consistent with QE results. 

%-------Simulations, foregrounds and systematics
\doparagraph{Robustness and consistency tests}
We test the consistency between $\Delta \hat D_{\ell_b}^{\rm BB}$ of data and simulations using templates built with different tracers. We subtract from $\Delta \hat D_{\ell_b}^{\rm BB}$ measured on the data the average of the same quantity computed with the non-Gaussian simulations $\av{\Delta \hat D_{\ell_b}^{\rm BB}}$ and fit to these band powers the amplitude parameter $\dabb$ of the fiducial binned $-\cbb$. By construction, $\dabb=0$ indicates a delensed $B$-power consistent with simulation expectations.  
We also build $
\chi^2$'s from $\Delta \hat D_{\ell_b}^{\rm BB}$ as follows: with $\Sigma_{bb'}$ the covariance of $\ddbb$ computed from the non-Gaussian simulations, we compute the data $\chi^2$ across all multipole bins $b$,
\ifwordcount
\else
\begin{equation}
	\chi^2\!\equiv\!\sum_{b,b'}\!\left(\!\Delta \hat D_{\ell_b}^{\rm BB}\!-\!\av{\!\Delta \hat D_{\ell_b}^{\rm BB}\!}\!\right)\!\Sigma^{-1}_{bb'}\!\left(\!\Delta \hat D_{\ell_{b'}}^{\rm BB}\!-\!\av{\!\Delta \hat D_{\ell_{b'}}^{\rm BB}\!}\!\right),
	\label{eq:chi2}
\end{equation}
\fi
that we turn into probability-to-exceed (PTE) values from the empirical ranking of the data $\chi^2$ compared to the results obtained for the simulations. Part of the noise and cosmic variance cancels in $\ddbb$, and this spectral difference is constrained about 4 times better (empirically) than the band powers themselves. 
In addition to $\hat \phi^{\rm QE}$, $\hat \phi^{\rm MAP}$, and $\hat \phi^{\rm MAP} - \hat \phi^{\rm QE}$  tracers, we used $\hat \phi^{\rm QE}$ removing modes  $L>500$ ($\hat \phi^{\rm QE,lowpass}$) to assess the impact of unmodeled tracer noise. To test for delensing bias we used a QE tracer $\hat \phi^{\rm QE, noOBD}$ built without OBD. 
Furthermore, we used tracers uncorrelated or anticorrelated with LSS, such as the lensing curl mode estimate $\hat \omega$~\cite{Pratten:2016dsm,Fabbian:2017wfp,Marozzi:2016qxl} (expected to be pure noise at our noise levels), $-\hat \phi^{\rm QE}$, and a QE tracer $\hat \phi^{\rm QE, \rm indep}$ estimated from an independent  simulation. This is independent from the map to delens, but has otherwise the same statistical properties. All these should produce no delensing and an increase of B-power after template subtraction.

\begin{table*}[htbp]
\ifwordcount
\else
\begin{ruledtabular}
\begin{tabular}{l|ccc|ccc}
$\Delta \hat D^{\rm BB}_{\ell_b} - \langle\Delta\hat  D_{\ell_b}^{\rm BB}\rangle $  & \patchun $\dabb$& \patchdeux $\dabb$& \patchtrois $\dabb$& \patchun PTE& \patchdeux PTE& \patchtrois PTE\\
\hline
$\hat \phi^{\rm QE}$ & $\phantom{-}0.01\pm0.12$ (0.13)$\phantom{-}$& $\phantom{-}0.09\pm0.13$ (0.10)$\phantom{-}$& $-0.02\pm0.15$ (0.05)$\phantom{-}$& 4\% & 60\% & 58\% \\ 
$\hat \phi^{\rm MAP}$ & $\phantom{-}0.07\pm0.12$ (0.15)$\phantom{-}$& $\phantom{-}0.04\pm0.14$ (0.11)$\phantom{-}$& $-0.05\pm 0.15$ (0.06)$\phantom{-}$& 4\%  & 84\%  & 75\% \\ 
$\hat \phi^{\rm QE, lowpass}$ & $-0.01\pm0.10$ (0.13)$\phantom{-}$& $\phantom{-}0.08\pm0.12$ (0.10)$\phantom{-}$& $-0.02\pm 0.13$  (0.07)$\phantom{-}$& 16\%  & 39\%  & 77\%\\ 
$\hat \phi^{\rm MAP}-\hat \phi^{\rm QE}$ & $\phantom{-}0.04\pm0.04$ (0.01)$\phantom{-}$& $-0.07\pm0.04$ (0.02)$\phantom{-}$ & $-0.04\pm0.04$ (0.02)$\phantom{-}$& 47\%  & 17\%  & 70\%\\ 
$\hat \phi^{\rm QE, noOBD}$ & $-0.01\pm0.16$ ($1.10$)$\phantom{-}$& $\phantom{-}0.01\pm0.19$ (1.09)$\phantom{-}$ & $-0.20\pm0.22$ (1.04)$\phantom{-}$& 14\%  & 80\% & 76\% \\ 
$\hat \omega^{\rm QE}$ & $-0.18\pm0.12$ ($-0.18$)& $-0.11\pm0.13$ ($-0.14$)& $\phantom{-}0.22\pm0.13$ ($-0.10$)& 13\%  & 3\%  & 22\% \\
-$\hat \phi^{\rm QE}$ & $\phantom{-}0.05\pm0.19$ ($-0.42$)& $-0.29\pm0.18$ ($-0.29$)& $\phantom{-}0.07\pm0.17$ ($-0.19$)& 7\%  & 14\%  & 46\% \\
$\hat \phi^{\rm QE, indep}$ & $\phantom{-}0.06\pm0.10$ ($-0.13$)& $\phantom{-}0.03\pm0.10$ ($-0.10$)& $-0.05\pm0.11$  ($-0.06$)& 62\%  & 5\%  & 98\% \\ 
\end{tabular}
\end{ruledtabular}
\fi
\caption{\label{table:chi2stats} 
Consistency tests between delensing observed on data $\Delta \hat D^{\rm BB}_{\ell_b}$ and simulations expectations $\langle\Delta\hat  D_{\ell_b}^{\rm BB}\rangle$ for different lensing tracers. For each patch we show the results of the fit of the amplitude $\dabb$ of our fiducial $-\cbb$ to $\Delta \hat D^{\rm BB}_{\ell_b} - \langle\Delta\hat  D_{\ell_b}^{\rm BB}\rangle$, as well as the $\chi^2$ PTEs for the consistency of such quantity with a null power spectrum. $\dabb$ fitted to $\langle\Delta\hat  D_{\ell_b}^{\rm BB}\rangle$ (which includes the delensing bias) is shown in parentheses. $\dabb>0$ means a reduction of $B$-power.
}
\end{table*}
Table~\ref{table:chi2stats} shows the summary of our tests. $\dabb$ amplitudes show no visible bias with respect to our simulations but we observe PTE values below 5\%, notably in \patchun. As all these tests are correlated we assessed the significance of these low PTEs simulating 20{,}000 realizations of all the band powers included in our test suites starting from their empirical covariance matrix estimated from our non-Gaussian simulations, and repeating the $\chi^2$ analysis. We found that the probability of observing three PTEs lower than 4\% in our test suite is 11\% and thus concluded that our data's low PTEs are not significant.

%-------Foreground and systematics
\doparagraph{Galactic foregrounds and systematics}
Polarized dust emission could affect delensing,  for example by adding Gaussian power to the tracer noise, and hence reducing the delensing efficiency. Since the dust angular power spectrum falls sharply with multipole $\ell$ and we use only $\ell_B \ge 500$, we expect this effect to be small.  The lensing estimator could also capture specific trispectra signatures in the highly non-Gaussian dust emission, which would propagate in lensing reconstruction and, later, delensing if uncorrected for. Preliminary studies suggest that at 150 GHz this effect is not important~\citep{Challinor:2017eqy}. It is implausible for such a signature to match the LSS deflection field; so this would also act to reduce the delensing efficiency. We quantified the expected impact of small-scale Gaussian polarized dust emission in our measurement by adding to our simulated datasets a template of this emission at our frequency produced with  Model 1 of the PySM package~\cite{pysm}, itself based on Planck \commander templates~\cite{planck2015fg}. Comparing simulated $\ddbb$ with and without dust we found a bias smaller than 1\% of the statistical error in all multipole bins. We ignored polarized galactic synchrotron contamination as it is subdominant in PB17~\cite{Ade:2017uvt}. Instrumental systematics effects in the \pb measurements of $\cbb$ and QE reconstruction were found to be negligible with respect to statistical uncertainties~\cite{Ade:2017uvt,spp-lensing}.

%-------Conclusions
\doparagraph{Conclusions}
Our analysis has achieved the highest internal $B$-mode delensing efficiencies to date, and is the first where the lensing tracer has been built from CMB polarization alone, serving as a proof of concept for future experiments where CMB polarization rather than temperature power will dominate the lensing tracer sensitivity. 
This work provides the first demonstration on deep polarization data that superior delensing efficiencies can indeed be achieved using iterative delensing methods~\cite{Hirata:2002jy, Carron:2017mqf}. This is a crucial step toward an efficient exploitation of future high-sensitivity $B$-mode polarization experiments of the next decade~\cite{Abazajian:2016yjj,Ade:2018sbj,litebird}, for which iterative methods will provide close-to-optimal constraints on the physics of inflation~\cite{Carron:2018lcr}.\\

%-------acknowledgements
\ifwordcount
\else
\begin{acknowledgements}
We thank Antony Lewis for discussion and comments.\*
The POLARBEAR project is funded by the National Science Foundation under Grants No.~AST-0618398 and No.~AST-1212230. The James Ax Observatory operates in the Parque Astron\'omico Atacama in Northern Chile under the auspices of the Comisi\'on Nacional de Investigaci\'on Cient\'ifica y Tecnol\'ogica de Chile (CONICYT).
JC and GF are supported by the European Research Council under the European Union's Seventh Framework Programme (FP/2007-2013) / ERC Grant Agreement No. [616170]. GF also acknowledges the support of the UK STFC grant ST/P000525/1.
BDS acknowledges support from an STFC Ernest Rutherford Fellowship. This work was supported by the World Premier International Research Center Initiative (WPI), MEXT, Japan. YC acknowledges the support from the JSPS KAKENHI Grants No. 18K13558, No. 18H04347, and No. 19H00674. The Melbourne group acknowledges support from the University of Melbourne and an Australian Research Council’s Future Fellowship (FT150100074). The SISSA group acknowledges support from the ASI-COSMOS network\footnote{\url{www.cosmosnet.it}} and the INDARK INFN Initiative\footnote{\url{web.infn.it/CSN4/IS/Linea5/InDark}}. The analysis presented here was also supported by the Moore Foundation Grant No. 4633, the Simons Foundation Grant No. 034079, and the Templeton Foundation Grant No. 58724. MH acknowledges the support from the JSPS KAKENHI Grants No. JP26220709 and No. JP15H05891. HN acknowledges the support from the JSPS KAKENHI Grant No. JP26800125. Support from the Ax Center for Experimental Cosmology at UC San Diego is gratefully acknowledged.  The APC group acknowledges travel support from Labex UNIVEARTHS. MAOAF acknowledges support from CONICYT UC Berkeley-Chile Seed Grant (CLAS fund) No. 77047, Fondecyt project 1130777 and 1171811, DFI postgraduate scholarship program and DFI Postgraduate Competitive Fund for Support in the Attendance to Scientific Events. NK acknowledges the support from JSPS Core-to-Core Program (A. Advanced Research Networks). AK acknowledges the support by JSPS Leading Initiative for Excellent Young Researchers (LEADER) and by the JSPS KAKENHI Grants No. JP16K21744 and No. JP18H05539. Work at LBNL is supported in part by the U.S. Department of Energy, Office of Science, Office of High Energy Physics, under Contract No. DE-AC02-05CH11231.
This research used resources of the National Energy Research Scientific Computing Center, a DOE Office of Science User Facility supported by the Office of Science of the U.S. Department of Energy under Contract No. DE-AC02-05CH11231 as well as resources of the Central Computing System, owned and operated by the Computing Research Center at KEK.
We acknowledge the use of the PySM\footnote{\url{https://github.com/bthorne93/PySM_public}}, LensIt\footnote{\url{https://github.com/carronj/LensIt}} and LensPix\footnote{\url{https://github.com/cmbant/lenspix}} packages.

\end{acknowledgements}
\bibliography{lensingbib}
\fi
\end{document}